\documentclass{jpsj2}
\usepackage{bm}

\title{First-Principles Momentum Dependent Local Ansatz Approach to the
Ground-State Properties of Iron-Group Transition Metals}

\author{Yoshiro Kakehashi\thanks{yok@sci.u-ryukyu.ac.jp, to be published in J. Phys. Soc. Jpn. {\bf 85} (2016).} and Sumal Chandra}
\inst{Department of Physics and Earth Sciences, Faculty of Science,\\
University of the Ryukyus, \\
1 Senbaru, Nishihara, Okinawa, 903-0213, Japan} 

\abst{
The ground-state properties of iron-group transition metals from Sc to 
Cu have been investigated on the basis of the first-principles momentum
dependent local ansatz (MLA) theory.  Correlation energy gain
is found to show large values for Mn and Fe: 0.090 Ry (Mn) and 0.094 Ry (Fe).
The Hund-rule coupling energies are found to be 
3000 K (Fe), 1400 K (Co), and 300 K (Ni).  It is sugested that  
these values can resolve the inconsistency in magnetic energy 
between the density functional theory and the first-principles dynamical
coherent potential approximation theory at finite temperatures.   
Charge fluctuations are shown to be suppressed by the intra-orbital 
correlations and inter-orbital charge-charge correlations, so that they 
show nearly constant values from V to Fe: 1.57 (V and Cr), 1.52 (Mn), and
1.44 (Fe), which are roughly twice as large as those
obtained by the $d$ band model.
The amplitudes of local moments are 
enhanced by the intra-orbital and inter-orbital spin-spin
correlations and show large values for Mn and Fe: 2.87 (Mn) and 2.58 (Fe).  
These values are in good agreement with the experimental values 
estimated from the effective Bohr magneton number and the inner core 
photoemission data.  
}

\kword{ first-principles variational theory, momentum-dependent local
ansatz, iron-group transition metals, electron correlations, correlation
energy, charge fluctuations, amplitude of local moment }

\begin{document}
\maketitle

\section{Introduction}

The iron-group transition metals and compounds are well-known to show a 
variety of physical properties in cohesion~\cite{jfcm77}, 
magnetism~\cite{pfu95,pfu12}, and superconductivity~\cite{imada10}.  
To understand their properties, enormous number of band
structure calculations have been made over the last several decades.
The density functional theory (DFT) has played the central role in
these calculations~\cite{hohen64,kohn65,rmm12}.  
The DFT with use of the exchange-correlation
potentials in the local density approximation (LDA)~\cite{barth72} 
or the generalized gradient approximation (GGA)~\cite{perdew86} 
is known to describe
quantitatively the ground-state properties such as the stability of
structure~\cite{bagno89}, lattice parameter~\cite{morruzi78}, 
bulk modulus~\cite{janak76}, as well as the magnetization~\cite{morruzi78}
and susceptibility~\cite{janak77} 
in many transition metals and compounds.

Although the DFT band theory is successful in many cases in spite of 
its simplicity, it is also known that 
the quantitativity of the DFT decreases with increasing
Coulomb interaction strength and the range of application is 
also limited in some cases.
In fact, the DFT is based on the Hohenberg-Kohn theorem~\cite{hohen64} 
which states
that the ground-state is given by the functional of electron density.
Thus the physical quantities expressed by the two-particle operators as
well as excitation spectra cannot be calculated by means of the DFT.  
Second, the DFT is based on the Kohn-Sham scheme~\cite{kohn65} 
in which the charge and spin 
densities are obtained from an independent electron system.  
Thus the momentum distribution
function as well as related mass enhancement factor cannot be described
by the DFT when the electron correlations become significant.
Furthermore, the LDA and GGA potentials in the DFT do not 
describe the orbital correlations as well as the Hund-rule correlations 
in the paramagnetic state~\cite{pfu95,pfu12}, 
and thus the ground-state energy is
overestimated in general in the paramagnetic state.

Because of the reasons mentioned above, the ground-state properties 
such as the correlation energy, the charge fluctuations, 
the amplitude of local moment, as well as the momentum distribution 
function have not yet been fully understood
from the quantitative point of view.  
In order to understand these ground-state properties, we have to take
alternative approaches such as the wavefunction 
method~\cite{mcg63,mcg64,bune00,bune12,schick12,yk14} 
and the dynamical mean
field theory (DMFT)~\cite{kotliar06,anisimov10} 
or equivalently the dynamical coherent potential
approximation (DCPA)~\cite{kake08-1,kake10,kake11,kake13}.

Early calculations of the ground-state properties of iron-group
transition metals have been made by Stollhoff and 
Fulde~\cite{stoll81,oles84,capell86} on the basis of
the local ansatz approach (LA) and the $d$ band model.  The LA is a
Gutzwiller-type variational approach in which the Hilbert space is
expanded by the residual Coulomb interactions to describe the local
electron correlations~\cite{gs77,gs78,gs80}.  
They obtained strong suppression of
charge fluctuations by a factor of two 
as well as strong enhancement of the amplitudes of LM
over 3$d$-series elements.  Since their calculations are based 
on the $d$ band
model and the LA does not reduce to the correct second order
perturbation theory in the weak Coulomb interaction limit, 
quantitative calculations are desirable to draw a solid conclusion.

Recently we proposed the momentum-dependent local ansatz (MLA) theory in
order to describe the ground-state properties 
quantitatively~\cite{kake08,pat11,pat13}.
The MLA improves the LA by taking into account all the two-particle
excited states with momentum-dependent amplitudes.  
In particular the theory reduces to the Rayleigh-Schr\"{o}dinger
perturbation theory in the weak Coulomb interaction limit as it should be, 
and describes well the correlated electrons from the
weak to strong Coulomb interaction regime.
In the next papers~\cite{chandra16,kake16}, 
we extended the MLA to the first-principle version
using the tight-binding LDA+U Hamiltonian~\cite{anisimov10} 
to describe quantitatively the ground-state
properties of the real systems and discussed electron correlation
effects in bcc Fe.

Alternative approach to quantitative description of correlated electrons 
is the first-principles DMFT (DCPA) combined with the LDA+U 
Hamiltonian~\cite{kotliar06,anisimov10,kake08-1,kake10,kake11,kake13}. 
The LDA+DMFT is a powerful method to strongly correlated electrons and 
has been applied to many systems.  The accuracy of the DMFT however 
strongly depends on the solver of the impurity problem for correlated 
electrons.  The Quantum Monte-Carlo method (QMC) can describe accurately 
the finite-temperature properties of the system.  But its efficiency 
is strongly reduced at low temperatures, and the QMC even causes the 
negative sign problem which prevents us from systematic investigations 
over wide range of interaction parameters.  The exact diagonalization 
method (ED) is useful to study exactly the physical properties at zero 
temperature.  But it cannot describe the low energy properties 
associated with the Fermi surface.  The numerical renormalization group 
theory (NRG) describes accurately the low energy excitations, but 
it does not accurately describe the excitations in high-energy region 
as well as the energy-integrated quantities.  Furthermore it is not 
applicable to the realistic systems because of the numerical difficulty.

The MLA on the other hand describes quantitatively the 
quasi-particle weight associated with the low energy excitations 
as well as the energy-integrated quantities 
such as the total energy and momentum distribution function without
numerical difficulty.  
In particular, we have shown in the recent paper~\cite{kake16} 
that the first-principles MLA quantitatively explains 
the mass enhancement factor of bcc Fe obtained by the ARPES 
experiment, while the LDA+DMFT combined with the three-body theory at 
zero temperature does not~\cite{sanchez09}.  
Furthermore the MLA allows us to calculate 
any static physical quantity because the wavefunction is known.  
These facts indicate that the first-principles MLA is competitive to 
the LDA+DMFT at zero temperature and thus it is a suitable approach to 
correlated electrons.

In the present paper, we investigate quantitatively the systematic 
change of the ground-state properties of iron-group transition metals 
from Sc to Cu on the basis of the first-principles MLA.
We present the correlation energy, the charge fluctuations, the
amplitude of local moment (LM) of the iron-group transition metals 
in the paramagnetic state.
We will clarify the systematic change of these quantities as a function of
the conduction electron number and elucidate the interplay among
intra-orbital correlations, the inter-orbital charge-charge correlations,
and the inter-orbital spin-spin correlations (, {\it i.e.}, the Hund-rule
correlations) in the ground-state properties.

In the following section, we outline the first-principles MLA.
Starting from the tight-binding LDA+U Hamiltonian and the local
ansatz wavefunction with momentum-dependent variational parameters, 
we obtain the ground-state energy in the single-site approximation
(SSA).  Using the variational principle, we derive the self-consistent 
equations for the variational parameters as well as related physical
quantities. 
In \S 3, we show the numerical results of calculations for the
correlation energy, the Hund-rule coupling energy, charge fluctuations,
and the amplitude of LM in the paramagnetic state.  We discuss
the Hund-rule coupling energy missing in the DFT+LDA band theory and 
its relation to the Curie temperatures in Fe and Co.  We
analyze the results with use of the three types of correlations, and
demonstrate that the suppression of charge fluctuations are dominated 
by the intra-orbital correlations and the inter-orbital charge-charge
correlations, while the enhancement of amplitude of LM is caused by the
intra-orbital correlations and the inter-orbital spin-spin
correlations.  In the last section, we summalize the results and future
problems.

\section{First-Principles MLA}

We adopt the first-principles LDA+U Hamiltonian with an atom 
in the unit cell~\cite{anisimov10, kake13}. 
\begin{align}
H & = \sum_{iL\sigma}\epsilon^{0}_{L} \ n_{iL\sigma} 
+ \sum_{iLjL^{'}\sigma}{t}_{iLjL^{'}}\ 
a^{\dagger}_{iL\sigma}\,{a}_{jL^{'}\sigma}  \nonumber \\
  & \hspace*{-4mm} + \sum_{i}\Big[ 
\sum_{m} {U}_{mm} n_{ilm\uparrow}\, n_{ilm\downarrow} 
+ \! \sum_{(m,m')} \! \Big(U_{mm'}-\frac{1}{2}J_{mm'}\Big)\, 
n_{ilm} n_{ilm{'}} - 
\!\! 2 \sum_{(m,m')} J_{mm'}\,
\bm{s}_{ilm} \! \cdot \! \bm{s}_{ilm'} \Big] \,.
\label{eqhldau}
\end{align}
Here $\epsilon^{0}_{L}$ is an atomic level of orbital $ L$ on site $i$. 
${t}_{iLjL'}$  is a transfer integral between $iL$ and $jL'$,  
$L=(l, m)$ being the $s \,(l=0)$, $p\, (l=1)$, and $d \,(l=2)$ 
orbitals~\cite{ander84,ander85}. 
$a^{\dagger}_{iL\sigma}{({a}_{iL\sigma})}$ is the creation (annihilation) 
operator for an electron on site $i$ with orbital $L$ and spin 
${\sigma}$, and 
$n_{iL\sigma} = a^{\dagger}_{iL\sigma}{a}_{iL\sigma}$ is 
the number operator on the same site $i$ with orbital $L$ and 
spin $\sigma$.  
The atomic level $\epsilon^{0}_{L}$ is calculated from the LDA atomic 
level $\epsilon_{L}$ by subtracting the $d$-$d$ Coulomb potential 
contribution~\cite{anisimov10,kake10}.
The third term at the rhs (right-hand-side) of Eq. (\ref{eqhldau})
denotes the on-site Coulomb interactions between $d$ electrons.
$U_{mm}\,(U_{mm'} )$ and $J_{mm'}$ are the intra-orbital (inter-orbital) 
Coulomb and exchange interactions between $d$ electrons, respectively. 
$n_{ilm}\, (\bm{s}_{ilm})$ with $l = 2 $ is the charge 
(spin) density operator for $d$ electrons on site $i$ and orbital $m$. 
The operator  $\bm{s}_{{iL}}$ is defined as 
$\bm{s}_{iL} = \sum_{\gamma \gamma'} a^{\dagger}_{iL\gamma} 
(\bm{\sigma})_{\gamma\gamma'}\, {a}_{iL\gamma'}/2$, $\bm{\sigma}$ 
being the Pauli spin matrices.

In the first-principles MLA, we split the Hamiltonian $H$ into 
the Hartree-Fock part $H_{0}$ and the residual interaction part 
$H_{\mathrm {I}}$: 
\begin{equation}
H = H_{0} + H_{\mathrm{I}} \,.
\label{eqhhfi}
\end{equation}
The latter is expressed as follows.
\begin{align}
H_{\mathrm{I}} &= \sum_{i} \Big[ \sum_{L} U_{LL}^{(0)}\ {O}^{(0)}_{iLL} 
+ \sum_{(L,\, L')} U_{LL'}^{(1)} \ {O}^{(1)}_{iLL'} + 
\sum_{(L,\, L')} U_{LL'}^{(2)}\ {O}^{(2)}_{iLL'} \Big] \,.  
\label{eqhi}
\end{align}
The first term is the intra-orbital Coulomb interactions, 
the second term is the inter-orbital charge-charge interactions, 
and the third term denotes the inter-orbital spin-spin interactions, 
respectively. 
The Coulomb interaction energy parameters $U_{LL'}^{(\alpha)}$ are 
defined by $U_{LL}\delta_{LL'}$ $(\alpha=0)$, 
$U_{LL'}-J_{LL'}/2$ $(\alpha=1) $, 
and $-2J_{LL'}$ $ (\alpha=2)$, respectively. 
The operators ${O}^{(0)}_{iLL}$, ${O}^{(1)}_{iLL'}$, 
and ${O}^{(2)}_{iLL'}$  are defined by
\begin{equation}
{O}^{(\alpha)}_{iLL^{\prime}} = 
\begin{cases}
		 \ \delta n_{ilm\uparrow} \, 
\delta n_{ilm\downarrow} \, \delta_{LL^{\prime}} &  \ (\alpha=0) \\
\ \delta n_{ilm} \, \delta n_{ilm'}  &  \ (\alpha=1) \\ 
\ \delta \bm{s}_{ilm} \cdot \delta \bm{s}_{ilm'} & \ (\alpha=2)\, .
\end{cases}
\label{eqoalph}
\end{equation}
Note that $\delta A$ for an operator $A$ is defined by 
$\delta A = A-\langle A\rangle_{0}$, $\langle \sim \rangle_{0}$ being 
the average in the Hartree-Fock approximation. 

When the Hamiltonian $H$ is applied to the Hartree-Fock wavefunction
$|\phi \rangle$, the Hilbert space is expanded by the local operators 
$\{ O^{(\alpha)}_{iLL^{\prime}} \}$ in the interactions.
In order to take into account these states as well as the states
produced in the weak Coulomb interaction limit, we introduce the 
momentum-dependent local correlators 
$\lbrace\tilde{O}^{(\alpha)}_{iLL'}\rbrace$ ($\alpha=$ 0, 1, and 2) 
as follows.
\begin{align}
\tilde{O}^{(\alpha)}_{iLL'}&= \sum_{\{kn\sigma\}}\langle{k'_{2}n'_{2}\vert iL}\rangle_{\sigma'_{2}} \langle{iL\vert {k}_{2}{n}_{2}}\rangle_{\sigma_{2}} \langle{k'_{1}n'_{1}\vert iL'}\rangle_{\sigma'_{1}} \langle{iL'\vert {k}_{1}{n}_{1}}\rangle_{\sigma_{1}}  \nonumber \\
&\hspace{1cm}\times\lambda^{(\alpha)}_{{LL'}\{{2'2 1'1}\}}\ \delta(a^{\dagger}_{k'_{2}n'_{2}\sigma'_{2}}a_{{k}_{2}{n}_{2}\sigma_{2}})\ \delta(a^{\dagger}_{k'_{1}n'_{1}\sigma'_{1}}a_{{k}_{1}{n}_{1}\sigma_{1}})\,.
\label{eqotilde}
\end{align}
Here $a^{\dagger}_{k n \sigma }{(a_{k n\sigma})}$ is the creation 
(annihilation) operator for an electron with momentum $\bm{k}$, 
band index $n$, and spin $\sigma $. These operators are given by those 
in the site representation as 
$a_{k n\sigma}=\sum_{iL}a_{iL\sigma}\langle k n\vert iL\rangle_{\sigma}$\,. 
$\langle k n\vert iL\rangle_{\sigma}$ are the overlap integrals
between the Hartree-Fock Bloch state $(\bm{k}n)$ and 
the local-orbital state $(iL)$.

The momentum-dependent parameters 
$\lambda^{(\alpha)}_{{LL'}\{{2'2 1'1}\}}$ in Eq. (\ref{eqotilde}) are
defined as
\begin{equation}
\lambda^{(0)}_{{LL'}\{{2'2 1'1}\}} = 
\eta_{L [2'2 1'1]}
\ \delta_{LL'}\,\delta_{\sigma'_{2}\downarrow}
\,\delta_{\sigma_{2}\downarrow}\,\delta_{\sigma'_{1}\uparrow}
\,\delta_{\sigma_{1}\uparrow}\,,
\label{eqlam0}
\end{equation}
\begin{equation}
\hspace{-1.5cm}\lambda^{(1)}_{{LL'}\{2'2 1'1\}} = 
\zeta^{(\sigma_{2}\sigma_{1})}_{LL' [2'2 1'1]}
\ \delta_{\sigma'_{2}\sigma_{2}}\, \delta_{\sigma'_{1}\sigma_{1}}\,,
\label{eqlam1}
\end{equation}
\begin{align}
\lambda^{(2)}_{{LL'}\{{2'2 1'1}\}} & = 
\sum_{\sigma}\xi^{(\sigma)}_{LL' [2'2 1'1]}
\ \delta_{\sigma'_{2}-\sigma}\  \delta_{\sigma_{2}\sigma}
\ \delta_{\sigma'_{1}\sigma}\ \delta_{\sigma_{1}-\sigma} \nonumber \\
&\hspace{1cm} + \frac{1}{2} \sigma_{1} \sigma_{2} 
\ \xi^{(\sigma_{2}\sigma_{1})}_{LL' [2'2 1'1]}
\ \delta_{\sigma'_{2}\sigma_{2}} \,\delta_{\sigma'_{1}\sigma_{1}}\,.                    
\label{eqlam2}
\end{align}
Here $\{2'21'1\}$($[2'21'1]$) implies that 
$\{2'21'1\} = k'_{2}n'_{2}\sigma'_{2}k_{2}n_{2}\sigma_{2} 
k'_{1}n'_{1}\sigma'_{1}k_{1}n_{1}\sigma_{1}$
( $[2'21'1] = k'_{2}n'_{2}k_{2}n_{2}k'_{1}n'_{1}k_{1}n_{1}$). 
$\eta_{L [2'21'1]}$, $\zeta^{(\sigma_{2}\sigma_{1})}_{L L'[2'21'1]}$, 
$\xi^{(\sigma)}_{L L'[2'21'1]}$, and 
$\xi^{(\sigma_{2}\sigma_{1})}_{L L'[2'21'1]}$ are the momentum-dependent
variational parameters.
It should be noted that 
$\tilde{O}^{(0)}_{iLL}$, $\tilde{O}^{(1)}_{iLL'}$, and 
$\tilde{O}^{(2)}_{iLL'}$ reduce to the local correlators, 
${O}^{(0)}_{iLL}$, ${O}^{(1)}_{iLL'}$, and ${O}^{(2)}_{iLL'}$ when 
$\eta_{L [2'21'1]} = \zeta^{(\sigma_{2}\sigma_{1})}_{L L'[2'21'1]}=1$ and 
$\xi^{(\sigma)}_{L L'[2'21'1]} = 
\xi^{(\sigma_{2}\sigma_{1})}_{L L'[2'21'1]}=1/2$, so that 
$\{ \tilde{O}^{(2)}_{iLL'} \}$ describe the intra-orbital correlations, 
the inter-orbital charge-charge correlations, and the inter-orbital
spin-spin correlations (, $i.e.,$ the Hund-rule correlations),
respectively. 

Using the correlators $\lbrace\tilde{O}^{(\alpha)}_{iLL'}\rbrace$ and 
the Hartree-Fock ground-state wavefunction $\vert{\phi}\rangle$, 
we construct the first-principles MLA wavefunction as follows.
\begin{equation}
\vert{\Psi}_\mathrm{MLA}\rangle = {\Big[\prod_{i}{ \Big( 
1 - \sum_{L}{\tilde{O}}^{(0)}_{iLL} - 
\sum_{(L,L')}{\tilde{O}}^{(1)}_{iLL'} - 
\sum_{(L,L')}{\tilde{O}}^{(2)}_{iLL'} \Big) } \Big]} 
\ \vert{\phi} \rangle \,.
\label{eqmla}
\end{equation}

The ground-state energy $\langle H \rangle$ is given by

\begin{align}
\langle H \rangle = \langle H\rangle_{0} + N\epsilon_c \,.
\label{eqener}
\end{align}
Here $\langle H \rangle_{0}$ denotes the Hartree-Fock energy, $N$ is the
number of atoms in the system.  $\epsilon_c$ is the correlation energy 
per atom defined by 
$N\epsilon_c \equiv \langle \tilde{H} \rangle = 
\langle H \rangle-\langle H \rangle_{0}$. 
Note that 
$\tilde{H}\equiv H -\langle H \rangle_{0} = \tilde{H}_{0}+H_{I}$. 
$\langle \sim \rangle$ ($\langle \sim \rangle_{0}$) denotes the full 
(Hartree-Fock) average with respect to 
$\vert \Psi_\mathrm{MLA}\rangle$ ($\vert \phi \rangle$).
The correlation energy $\epsilon_c$ is expressed in the single-site 
approximation (SSA) as follows~\cite{chandra16}.
\begin{equation}
{\epsilon_c} = \frac{{- \langle {\tilde{O_i}^\dagger}} {H}_{I}\rangle_0 
- \langle {H}_{I} \tilde{O_i}\rangle_0 
+ \langle {\tilde{O_i}^\dagger} \tilde{H} \tilde{O_i} \rangle_0 }
{1 + \langle \tilde{O_i}^\dagger\tilde{O_i} \rangle_0} \,. 
\label{eqcorr}
\end{equation}
Here $\tilde{O_i} = \sum_{\alpha} \sum_{\langle L,\, L' \rangle} 
\tilde{O}^{(\alpha)}_{iLL'}$. 
The sum $\sum_{\langle L,\, L' \rangle}$ is defined by a single sum 
$\sum_{L}$ when  $L'$=$L$, and by a pair sum $\sum_{(L,\, L')}$ 
when $L' \neq L$.
Each element in Eq. (\ref{eqcorr}) has been calculated with use of 
Wick's theorem.

The variational parameters are determined from the stationary condition 
$\delta\epsilon_{c}=0$ as follows.
\begin{equation}
-\langle(\delta \tilde{O}_{i}^{\dagger}){H}_{I}\rangle_0 + 
\langle (\delta\tilde{O}_{i}^{\dagger} )\tilde{H} \tilde{O_i}\rangle_0 -
\epsilon_c {\langle(\delta\tilde{O}_{i}^{\dagger}) \tilde{O_i}\rangle_0}
+ c.c. = 0 \,.
\label{eqvar}
\end{equation}
Here $\delta \tilde{O}_{i}^{\dagger}$ denotes the variation of 
$\tilde{O}_{i}^{\dagger}$ with respect to 
$\{ \lambda^{(\alpha)}_{{LL'}\{2'2 1'1\}} \}$.

Since it is not easy to solve Eq. (\ref{eqvar}) for arbitrary Coulomb
interaction strength, we
make use of the following ansatz for the variational parameters, 
which interpolates between the weak Coulomb interaction limit and 
the atomic limit. 
\begin{equation}
\lambda^{(\alpha)}_{{LL'}\{{2'2 1'1}\}} = 
\frac{ U_{LL'}^{(\alpha)} \sum_{\tau} 
C_{\tau\sigma_{2}^{'}\sigma_{2}\sigma_{1}^{'}\sigma_{1}}^{(\alpha)}
\ \tilde{\lambda}_{\alpha\tau L L'}^{(\sigma_{2}\sigma_{1})} }
{ \epsilon_{k^{\prime}_{2} n^{\prime}_{2} \sigma^{\prime}_{2}} - 
\epsilon_{k_{2} n_{2} \sigma_{2}} - 
\epsilon_{k^{\prime}_{1} n^{\prime}_{1} \sigma^{\prime}_{1}} - 
\epsilon_{k_{1} n_{1} \sigma_{1}}
- \epsilon_{\rm c} } \,.
\label{eqvlam}
\end{equation}
Here the spin-dependent coefficients 
$C_{\tau\sigma_{2}^{'}\sigma_{2}\sigma_{1}^{'}\sigma_{1}}^{(\alpha)}$ 
are defined by 
$\delta_{\sigma'_{2}\downarrow}\,\delta_{\sigma_{2}\downarrow}\,
\delta_{\sigma'_{1}\uparrow}\,\delta_{\sigma_{1}\uparrow}$  
($\alpha=0$), 
$\delta_{\sigma'_{2}\sigma_{2}} \,\delta_{\sigma'_{1}\sigma_{1}}$  
($\alpha=1$), 
$-(1/4)\ \ \sigma_{1}\sigma_{2}\delta_{\sigma'_{2}\sigma_{2}} 
\delta_{\sigma'_{1}\sigma_{1}}$ ($\alpha=2,\tau=l$), and 
$-(1/2)\sum_{\sigma}\delta_{\sigma'_{2} - \sigma} 
\delta_{\sigma_{2}\sigma} \delta_{\sigma'_{1}\sigma} 
\delta_{\sigma_{1}-\sigma}$ ($\alpha=2,\tau=t$), respectively. 
Note that $l\,(t)$ implies the longitudinal (transverse) component. 
The parameters
$\tilde{\lambda}_{\alpha\tau L L'}^{(\sigma\sigma')}$ 
in Eq. (\ref{eqvlam}) are defined by 
$\tilde{\eta}_{LL'} \delta_{LL'}\delta_{\sigma'-\sigma}$ $(\alpha=0)$,  
$\tilde{\zeta}_{LL'}^{(\sigma\sigma')}$ $(\alpha=1)$, 
$\tilde{\xi}_{tLL'}^{(\sigma)}\delta_{\sigma'-\sigma}$ 
$(\alpha=2, \tau=t)$, and 
$\tilde{\xi}_{lLL'}^{(\sigma\sigma')}$ $(\alpha=2,\tau=l)$, 
respectively. 
The renormalization factors 
$\tilde{\eta}_{LL}$, $\tilde\zeta^{(\sigma\sigma')}_{LL'}$, 
$\tilde \xi^{(\sigma)}_{tLL'}$, and 
$\tilde \xi^{(\sigma\sigma')}_{lLL'}$ are new variational 
parameters to be determined.
The denominator in Eq. (\ref{eqvlam}) expresses the two-particle
excitation energy.  $\epsilon_{kn\sigma}$ denotes the Hartree-Fock one
electron energy eigenvalue for momentum $\bm{k}$, 
band index $n$, and spin $\sigma$.
Note that when 
$\tilde{\eta}_{LL} = \tilde\zeta^{(\sigma\sigma')}_{LL'} = 1$ 
and $\tilde \xi^{(\sigma\sigma')}_{lLL'} = 
\tilde \xi^{(\sigma)}_{tLL'} = -1$, the MLA wavefunction (\ref{eqmla})
reduces to that of the Rayleigh-Schr\"{o}dinger perturbation theory in
the weak Coulomb interaction limit.

Substituting Eq. (\ref{eqvlam}) into the elements in Eq. (\ref{eqvar}), 
we obtain the self-consistent equations for the variational parameters.
In the paramagnetic case, the variational parameters 
$\tilde{\lambda}_{\alpha\tau L L'}^{(\sigma\sigma')}$ are spin 
independent (, $i.e., \tilde{\lambda}_{\alpha\tau L L'} $), 
and the self-consistent equations are expressed as 
follows~\cite{chandra16}. 
\begin{equation}
\tilde{\lambda}_{\alpha\tau LL'} = 
\tilde{Q}_{LL'}^{-1} \left( \kappa_{\alpha} P_{LL'} - 
{U}^{(\alpha) -1}_{LL'} \, K^{(\alpha)}_{\tau LL'} \right) \, .
\label{eqsclam}
\end{equation}
Here $\tilde{Q}_{LL'}$ has the form 
$\tilde{Q}_{LL'}={Q}_{LL'}-{\epsilon_c} S_{LL'}$. 
The constant $\kappa_{\alpha}$ is defined by $1$ for $\alpha=0, 1$, and 
$-1$ for $\alpha=2$.
The second terms at the rhs of Eq. (\ref{eqsclam}) originates in the 
matrix element 
$\langle {\tilde{O_i}^\dagger} H_{I} \tilde{O_i} \rangle_0$, 
{\it i.e.}, the
third term in the numerator of the correlation energy (\ref{eqcorr}).
These terms are of higher order in Coulomb interaction and are given
by a linear combination of $\{ \tilde{\lambda}_{\alpha\tau LL'} \}$.
${Q}_{LL'}$, $S_{LL'}$, $P_{LL'}$, and $K^{(\alpha)}_{\tau LL'}$ are 
expressed by means of the Laplace transforms of the Hartree-Fock 
local densities of states~\cite{chandra16}.

It should be noted that $\tilde{Q}_{LL'}$, $P_{LL'}$, and 
$K^{(\alpha)}_{\tau LL'}$ 
contain the correlation energy $\epsilon_{c}$ and the Fermi level 
$\epsilon_{F}$.  Moreover $K^{(\alpha)}_{\tau LL'}$ are given by the
linear combination of $\{ \tilde{\lambda}_{\alpha\tau LL'} \}$.
The correlation energy $\epsilon_{c}$ is expressed by 
Eq. (\ref{eqcorr}) with variational parameters (\ref{eqvlam}). 
The Fermi level $\epsilon_{F}$ is determined by the 
conduction electron number per atom $n_{e}$, which is expressed as
\begin{equation}
n_{e}=\sum_{L}\langle n_{iL}\rangle\,.
\label{eqne}
\end{equation}
Taking the same steps as in Eq. (\ref{eqcorr}), we obtain the 
partial electron number of orbital $L$ on site $i$ in the SSA as follows. 
\begin{equation}
\langle n_{iL}\rangle = \langle n_{iL}\rangle_{0} + 
\langle \tilde{n}_{iL}\rangle \,.
\label{eqnil}
\end{equation}
Here $\langle n_{iL}\rangle_{0}$ denotes the  Hartree-Fock electron 
number. The correlation correction $\langle \tilde{n}_{iL}\rangle$ is 
expressed as follows. 
\begin{align}
\langle \tilde{n}_{iL}\rangle = \frac{ \langle 
\tilde{O}_{i}^{\dagger} \tilde{n}_{iL} \tilde{O}_{i} \rangle_{0}}
{1 + \langle \tilde{O_i}^{\dagger}\tilde{O_i} \rangle_0 } \, .
\label{eqnilcorr}
\end{align}
Note that $\langle \tilde{O}_{i}^{\dagger} \tilde{n}_{iL} \rangle_{0}$
and $\langle \tilde{n}_{iL} \tilde{O}_{i}^{\dagger} \rangle_{0}$, 
which correspond to the first and second terms in the numerator of the
correlation energy (\ref{eqcorr}), vanish according to Wick's theorem.
The other elements at the rhs of Eq. (\ref{eqnilcorr}) 
are also calculated by using Wick's theorem.
Equations (\ref{eqcorr}), (\ref{eqsclam}), and (\ref{eqne}) determine 
self-consistently the correlation energy $\epsilon_{c}$, 
the Fermi level $\epsilon_{F}$, 
as well as the variational parameters 
$\{\tilde{\lambda}_{\alpha\tau LL'}\}$. 

The local charge fluctuation and the amplitude of the local moment 
for $d$ electrons are calculated from the following expressions. 
\begin{align}
\langle (\delta n_{d})^{2}\rangle & = 
\sum_{L\sigma}^{d} \langle n_{iL\sigma} \rangle_{0} \ 
(1-\langle n_{iL\sigma}\rangle_{0}) + 
\sum_{L\sigma}^{d} \langle \tilde{n}_{iL\sigma} \rangle \ 
(1-2\langle n_{iL\sigma}\rangle_{0}) \nonumber \\
&\hspace{.5cm} - \langle\tilde{n}_{id} \rangle^{2} + 
2\sum_{L}^{d} \langle O_{iLL}^{(0)} \rangle + 
2\sum_{(L,L')}^{d} \langle O_{iLL'}^{(1)} \rangle \,,
\label{eqnd2}
\end{align}
\begin{align}
\langle \bm{S}^{2}\rangle & = 
\frac{3}{4}\sum_{L\sigma}^{d} \langle n_{iL\sigma} \rangle_{0} \ 
(1 - \langle n_{iL\sigma} \rangle_{0}) + 
\frac{3}{4} \sum_{L\sigma}^{d} \langle \tilde{n}_{iL\sigma} \rangle \ 
(1 - 2\langle n_{iL-\sigma} \rangle_{0}) \nonumber \\
&\hspace{.5cm} - \frac{3}{2}\sum_{L}^{d} \langle O_{iLL}^{(0)} \rangle 
+ 2\sum_{(L,L')}^{d} \langle O_{iLL'}^{(2)} \rangle \,.
\label{eqs2}
\end{align}
Here the first terms at the rhs of Eqs. (\ref{eqnd2}) and (\ref{eqs2}) 
express the Hartree-Fock contributions. 
$\langle\tilde{n}_{iL\sigma}\rangle$ in the second terms is given by 
Eq. (\ref{eqnilcorr}) in which $\tilde{n}_{iL}$ has been replaced by 
$\tilde{n}_{iL\sigma}$. 
$\langle\tilde{n}_{id}\rangle$ in the third term of Eq. (\ref{eqnd2}) 
is defined by $\sum_{L}^{d} \langle \tilde{n}_{iL} \rangle$. 
Note that $\sum_{L}^{d}$ denotes the sum over $d$ orbitals ($l=2$). 
The remaining correlation 
corrections at the rhs of Eqs. (\ref{eqnd2}) and (\ref{eqs2}) are 
obtained from the residual interaction elements 
$\langle O_{iLL'}^{(\alpha)}\rangle$, which are expressed in the SSA 
as follows.  
\begin{equation}
\sum_{\langle L,\, L' \rangle} \langle O_{iLL'}^{(\alpha)} \rangle = 
\frac{\displaystyle{- \sum_{\langle L,\, L' \rangle} \langle 
{\tilde{O_i}^\dagger}} {O}_{iLL'}^{(\alpha)} \rangle_0 - 
\sum_{\langle L,\, L' \rangle} \langle 
{O}_{iLL'}^{(\alpha)} \tilde{O_i} \rangle_0 
+ \sum_{\langle L,\, L' \rangle} \langle 
{\tilde{O_i}^\dagger} {O}_{iLL'}^{(\alpha)} 
\tilde{O_i} \rangle_0 }
{1 + \langle{\tilde{O_i}^\dagger\tilde{O_i}} \rangle_0 } \,. 
\label{eqoicorr}
\end{equation}
The elements at the rhs of Eq. (\ref{eqoicorr}) are again
calculated with use of Wick's theorem~\cite{chandra16}.

\section{Numerical Results}

\begin{figure}
\begin{center}
\includegraphics[width=10cm]{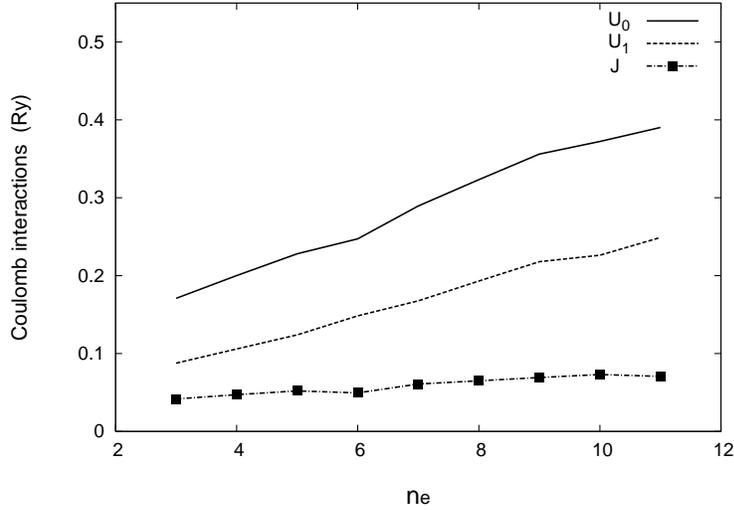}
\end{center}
\caption{
Intra-atomic Coulomb and exchange energy parameters as a function of 
the conduction electron number of iron-group transition metals.
These parameters are obtained from the
band~\cite{bdyo89} and atomic~\cite{mann67} calculations.  
Intra-orbital Coulomb interactions $U_{0}$: solid curve,
inter-orbital Coulomb interactions $U_{1}$: dashed curve,
exchange interactions $J$: closed squares and dot-dashed curve.
}
\label{figuj}
\end{figure}

In the present calculations, we adopted the same lattice constants and
structures as used by Andersen {\it et al.}~\cite{ander85}.
We constructed the tight-binding (TB) LDA+U Hamiltonians 
with use of the Barth-Hedin exchange-correlation potentials and the TB
linear muffin-tin orbital (LMTO) method within the atomic sphere
approximation (ASA).  Furthermore we adopted 
orbital-independent Coulomb and exchange interactions 
$U_{mm}=U_{0}$, $U_{mm'}=U_{1}$ ($m' \neq m$), and $J_{mm'}=J$. 
These values are obtained from the average Coulomb interaction 
energies $U$ via the relations $U_{0}=U+8J/5$ and $U_{1}=U-2J/5$, where
we used the relation $U_{0}=U_{1}+2J$ for the cubic system.
We applied the average interactions $U$ obtained by Bandyopadhyay 
{\it et al.}~\cite{bdyo89} 
and the average $J$ obtained from the Hartree-Fock 
atomic calculations~\cite{mann67}. 
The Coulomb and exchange interaction energies from Sc and Cu are 
depicted in Fig. \ref{figuj} as a function of the conduction 
electron number $n_{e}$.
The same Hamiltonians and Coulomb-exchange interactions have been 
applied in the investigations of the excitation spectra in 3$d$
transition metals with use of the first-principles dynamical 
CPA~\cite{kake10}.
%
%
%
%

\begin{figure}
\begin{center}
\includegraphics[width=10cm]{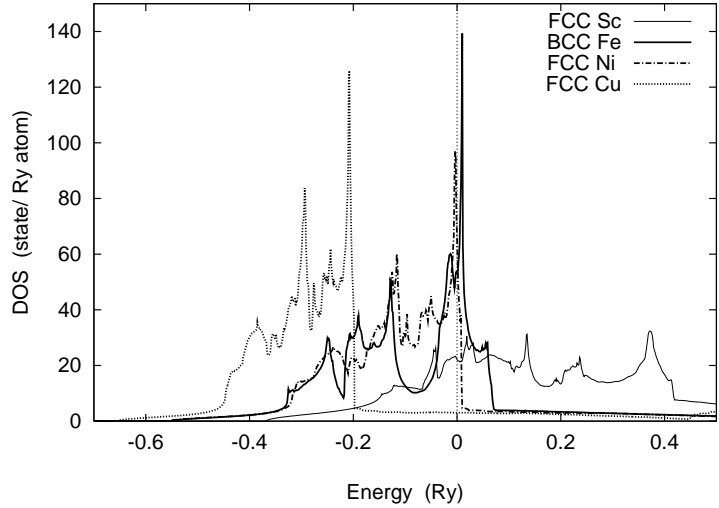}
\end{center}
\caption{
Calculated densities of states (DOS) for fcc Sc (thin solid curve),
 paramagnetic bcc Fe (solid curve), paramagnetic fcc Ni (dot-dashed
 curve), and fcc Cu (dotted curve) in the Hartree-Fock approximation.
}
\label{figdos}
\end{figure}

We performed the self-consistent Hartree-Fock calculations from Sc
to Cu in the paramagnetic state using the TB LDA+U Hartree-Fock
Hamiltonian $H_{0}$ (see Eq. (\ref{eqhhfi})).
Figure \ref{figdos} shows some total densities of states (DOS) obtained
by the self-consistent calculations.  The $d$ bands sink commonly by
$0.02 \sim 0.03$ Ry as compared with those in the LDA bands.
The $d$ band widths are broader than the LDA bands for the elements 
with $d$ electrons less than half, 
by 30 \% for fcc Sc, 15 \% for fcc Ti, 10 \% for bcc V, and 4
\% for bcc Cr, respectively.  The widths 
shrink for the elements with $d$ electrons more than half, by 0 \% for
fcc Mn, 5 \% for bcc Fe, fcc Co and fcc Ni, and 8 \% for fcc Cu, 
respectively, as compared with the LDA bands.
However the DOS below the Fermi level are basically the same as those in
the LDA bands except fcc Cu in which the $d$ bands are shifted by 0.086
Ry towards the lower energy as compared with the LDA.

\begin{figure}
\begin{center}
\includegraphics[width=10cm]{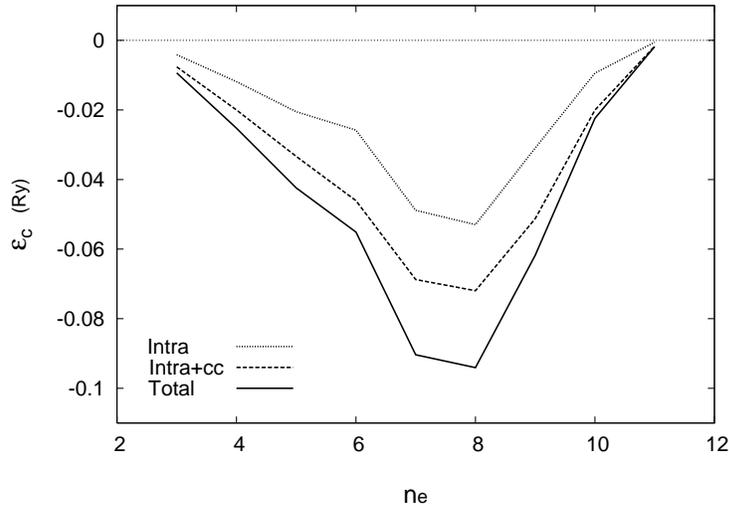}
\end{center}
\caption{
Systematic change of correlation energy from Sc to Cu as a function of
 conduction electron number $n_{e}$.  Dotted curve: correlation energy
 due to intra-orbital correlations $\epsilon_{c}({\rm intra})$, 
dashed curve: correlation energy due to intra-orbital and 
inter-orbital charge-charge correlations $\epsilon_{c}({\rm intra+cc})$, 
solid curve: total correlation
 energy including the inter-orbital spin-spin correlations 
$\epsilon_{c}({\rm total})$.
}
\label{figec}
\end{figure}

We solved the self-consistent equations (\ref{eqcorr}), 
(\ref{eqsclam}), and (\ref{eqne}) using the
Hartree-Fock energy bands and eigen vectors.
Figure \ref{figec} shows calculated correlation energies from Sc to Cu.
Correlation energy gain $|\epsilon_{c}|$ increases first 
with increasing the conduction electron number $n_{e}$, 
shows a large value 0.090 Ry for fcc Mn and the maximum value 
0.094 Ry for bcc Fe.  Then it rapidly decreases with further 
increasing $n_{e}$.  
The correlation energy gain is negligible for Cu since the $d$ 
electron states are almost occupied.  Detailed values of
correlation energies from Sc to Cu are presented in Table \ref{tblec}.
%
%
%
%

In order to clarify the role of three types of correlations 
introduced in the MLA wavefunction (\ref{eqmla}), we 
calculated the correlation energy due to the intra-orbital correlations
$\epsilon_{c}({\rm intra})$, the correlation energy due to the 
intra-orbital and inter-orbital charge-charge correlations 
$\epsilon_{c}({\rm intra+cc})$,
and the total correlation energy $\epsilon_{c}({\rm total})$ with the
inter-orbital spin-spin correlations (, {\it i.e.}, the Hund-rule
correlations).  They are defined by $\epsilon_{c}$ when 
$\tilde{\zeta}_{LL'}=\tilde{\xi}_{lLL'}=\tilde{\xi}_{tLL'}=0$,
$\epsilon_{c}$ when $\tilde{\xi}_{lLL'}=\tilde{\xi}_{tLL'}=0$, 
and the full $\epsilon_{c}$, respectively.
Note that $|\epsilon_{c}({\rm intra})|$ is the correlation energy 
gain due to the reduction of double occupancy on the same orbital.  
Therefore the energy gain is expected to show the maximum for the
half-filled $d$ bands.
The shape of the curve is similar to the full $\epsilon_{c}$ as shown 
in Fig. \ref{figec}.  The contribution of $\epsilon_{c}({\rm intra})$ 
to $\epsilon_{c}({\rm total})$ is about 50 \%.
The difference between $\epsilon_{c}({\rm intra+cc})$ and
$\epsilon_{c}({\rm intra})$ implies the energy gain due to 
the intra-orbital
charge-charge correlations.  It is significant in Sc, Ti, and Ni,
and makes about 50 \% contribution to $\epsilon_{c}({\rm total})$.
In the case of Mn and Fe, its contribution is about 25 \%.
 
The difference between $\epsilon_{c}({\rm total})$ and
$\epsilon_{c}({\rm intra+cc})$ indicates the energy gain 
due to the inter-orbital spin-spin correlations, {\it i.e.}, 
the Hund-rule correlations. 
The Hund-rule correlation energy becomes significant for Mn and Fe and
amounts to about 25 \% of $\epsilon_{c}({\rm total})$. 
The energy is small for fcc Ni, and is negligible for Cu.
In Table \ref{tblhund}, we summarize the Hund-rule coupling 
energies in iron-group transition metals, which are defined by 
$\Delta \epsilon_{\rm H} \equiv 
\epsilon_{c}({\rm intra+cc}) - \epsilon_{c}({\rm total})$. 
Mn and Fe show large Hund-rule energies: 0.29 eV (= 0.0216 Ry) and 0.30 eV
(= 0.0221 Ry), respectively, 
which are about 3000 K, indicating that these metals 
have a well-defined local magnetic moment above the Curie or
N\'{e}el temperature.  In the case of Ni, it is 0.032 eV (= 0.0024 Ry), 
which is roughly 300 K, indicating that the Hund-rule correlations do 
not play an important role above the Curie temperature (640 K) in the
case of Ni.  
%
%
%
%

The DFT+LDA band theory does not take into account the Hund-rule
coupling energy in the paramagnetic state, 
while the theory does the energy via the polarized exchange-correlation 
potential in the ferromagnetic state.  
Thus the magnetic energy $E_{\rm mag}$ defined by
the energy difference between the paramagnetic state and the
ferromagnetic state is in general overestimated in the DFT+LDA.  In
fact, the magnetic energy of the bcc Fe (fcc Co) in the DFT+LDA, {\it
i.e.}, $E_{\rm mag}({\rm LDA})$ is estimated to be about 5000 
K~\cite{gun76} (4000 K~\cite{staun92}), 
while the first-principles DCPA yields $E_{\rm mag}({\rm DCPA}) \sim 2000$ K
($2500$ K)~\cite{kake11}.  
The discrepancies in two approaches are qualitatively
explained by taking into account the Hund-rule coupling 
$\Delta \epsilon_{\rm H}$ obtained in the present calculation.  In fact,
if we subtract $\Delta \epsilon_{\rm H} \sim 3000$ K ($1400$ K) from 
$E_{\rm mag}({\rm LDA}) \sim 5000$ K ($4000$ K) for bcc Fe (fcc Co), we
find $E_{\rm mag} \sim 2000$ K ($2600$ K), which is comparable to 
$E_{\rm mag}({\rm DCPA}) \sim 2000$ K ($2600$ K) obtained by the 
first-principles DCPA. 

In the case of Ni, $E_{\rm mag}({\rm LDA}) \sim 2900$ K~\cite{staun92}, 
while $\Delta \epsilon_{\rm H} \sim 300$ K.  Therefore, the Hund-rule 
coupling energy does not explain a large difference between 
$E_{\rm mag}({\rm LDA})$ and $E_{\rm mag}({\rm DCPA}) \sim 600$ 
K~\cite{kake11}.  
The DFT+LDA band theory overestimates the exchange 
splitting in Ni by a factor of two~\cite{gun76,east78,marten84,lieb79}.  
It is possible that additional 
error in magnetic energy associated with the overestimate of 
the exchange splitting explains the discrepancy.  Spin-polarized MLA 
calculations are desired to clarify the origin of the discrepancy in Ni. 
%
%
\begin{table}
\caption{
Calculated correlation energies $\epsilon_{c}({\rm total})$ (Ry) 
for iron-group transition metals.  
}
\label{tblec}
\vspace*{5mm}
\begin{tabular}{ccccccccc}
\hline
Sc  & Ti  & V  & Cr  & Mn  & Fe  & Co  & Ni  & Cu \\ \hline 
$-0.0094$ & $-0.0253$ & $-0.0424$ & $-0.0551$ & $-0.0904$ & $-0.0941$ & 
$-0.0618$ & $-0.0224$ & $-0.0018$ \\ \hline
\end{tabular}
\end{table}
%
%
%
%
\begin{table}
\caption{
The Hund-rule coupling energies $\Delta \epsilon_{\rm H} \equiv
 \epsilon_{c}({\rm intra+cc}) - \epsilon_{c}({\rm total})$ (eV) 
in iron-group transition metals.  
}
\label{tblhund}
\vspace*{5mm}
\begin{tabular}{ccccccccc}
\hline
Sc  & Ti  & V  & Cr  & Mn  & Fe  & Co  & Ni  & Cu \\ \hline 
0.0240 & 0.0717 & 0.1229 & 0.1228 & 0.2942 & 0.3008 & 0.1432 & 0.0323 & 0.0012 \\ \hline
\end{tabular}
\end{table}
%
%

\begin{figure}
\begin{center}
\includegraphics[width=10cm]{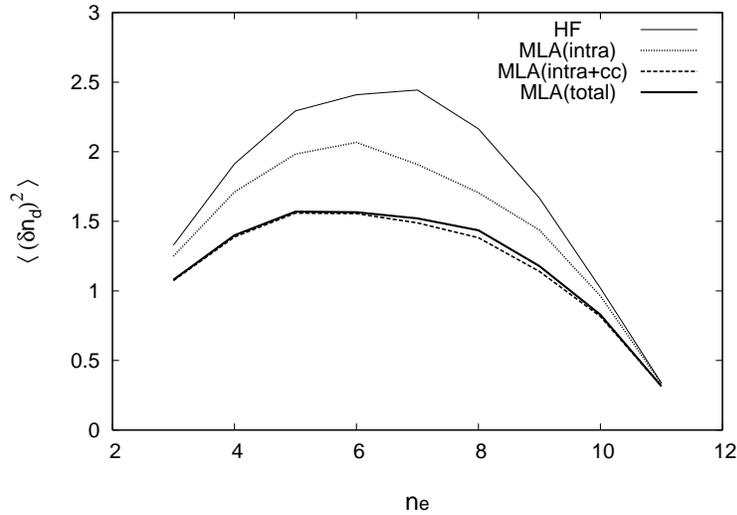}
\end{center}
\caption{
Systematic change of charge fluctuations 
$\langle (\delta n_{d})^{2} \rangle$ from Sc to Cu as a function of
 conduction electron number $n_{e}$.  Thin curve: charge fluctuations in
 the Hartree-Fock approximation 
$\langle (\delta n_{d})^{2} \rangle({\rm HF})$, 
dotted curve: charge fluctuations 
 with intra-orbital correlations 
$\langle (\delta n_{d})^{2}\rangle({\rm intra})$, dashed curve: 
charge fluctuations with intra-orbital and inter-orbital charge-charge 
correlations $\langle (\delta n_{d})^{2}\rangle({\rm intra+cc})$, 
solid curve: charge fluctuations
 with full correlations including the inter-orbital spin-spin correlations 
$\langle (\delta n_{d})^{2}\rangle({\rm total})$.
}
\label{figdn2}
\end{figure}

Charge fluctuations associated with electron hopping are suppressed by
electron correlations.  We calculated the local charge fluctuations of
$d$ electrons $\langle (\delta n_{d})^2\rangle$.  
As shown in Fig. \ref{figdn2}, the charge fluctuations for the
Hartree-Fock independent electrons show a parabolic behavior with
the maximum 2.44 at $n_{e}=7$ (Mn) as a function of the conduction
electron number $n_{e}$.  
However the charge fluctuations based
on the first-principles MLA are suppressed, and are 
approximately constant from V to Fe 
due to electron correlations: 1.57, 1.57, 1.52, and 1.44 for V, 
Cr, Mn, and Fe, respectively.

Early calculations based on the LA and the $d$ band 
model with common crystal structure (bcc), and common Coulomb and
exchange energy parameters ($U/W_{\rm d}=0.5$ and $J/U=0.2$, $W_{\rm d}$
being the $d$ band width)~\cite{capell86} 
show $\langle (\delta n_{d})^2\rangle = $ 0.85, 0.70, 0.68, and 0.87 for
the $d$ electron number $n_{d} = $ 4 (V), 5 (Cr), 6 (Mn), and 7 (Fe). 
These results are considerably 
smaller than those obtained by the first-principles MLA calculations, and 
overestimate the atomic character in charge fluctuations by a factor of 
two.  The LA approximation to the charge fluctuations introduces an error
typically by about 10 \% according to our previous 
calculations~\cite{kake08,pat11}. 
Although the LA + $d$-band model calculations mentioned above are not 
quantitative version, we suggest that the large discrepancy in charge 
fluctuations between the two theoretical calculations is mainly 
attributed to the neglect of 
the hybridization between the $sp$ and $d$ electrons in the model 
calculations.
%
%
%
%

We investigated the role of three kinds of electron correlations in
the charge fluctuations, taking the same steps as in the correlation
energy.
As shown in Fig. \ref{figdn2}, the intra-orbital correlations make a
significant contribution to the suppression of the charge fluctuations
(see the difference between
$\langle (\delta n_{d})^{2}\rangle({\rm intra})$ and 
$\langle (\delta n_{d})^{2}\rangle ({\rm HF})$).
In the case of Mn and Fe, their contributions amount to more than 50 \%
of the total suppression of the charge fluctuations 
(, {\it i.e.}, $\langle (\delta n_{d})^{2} \rangle({\rm HF}) - 
\langle (\delta n_{d})^{2} \rangle({\rm total})$).
The inter-orbital charge-charge correlations also make the contributions
being comparable to the intra-orbital ones (see the difference between
$\langle (\delta n_{d})^{2}\rangle({\rm intra+cc})$ and 
$\langle (\delta n_{d})^{2}\rangle({\rm intra})$). 
The contribution becomes significant when the $d$ electron number
deviates from the half filling.  On the other hand, the intra-orbital
spin-spin correlations (the Hund-rule correlations) hardly make 
contribution to
the charge fluctuations as shown in Fig. \ref{figdn2} 
(see the difference between 
$\langle (\delta n_{d})^{2}\rangle({\rm total})$ 
and $\langle (\delta n_{d})^{2}\rangle({\rm intra+cc})$).

\begin{figure}
\begin{center}
\includegraphics[width=10cm]{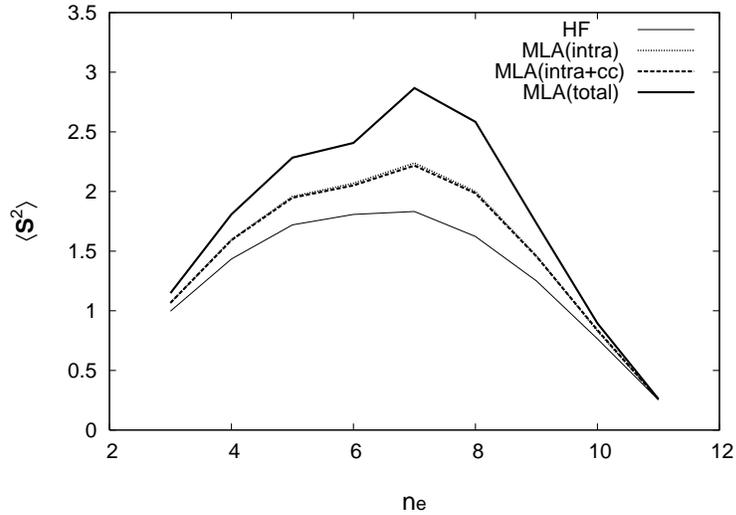}
\end{center}
\caption{
Systematic change of amplitude of local magnetic moment (LM) 
$\langle \bm{S}^2\rangle$ from Sc to Cu as a function of
 conduction electron number $n_{e}$.  Thin curve: amplitude of LM in
 the Hartree-Fock approximation, dotted curve: amplitude of LM 
 with the intra-orbital correlations 
$\langle \bm{S}^2 \rangle({\rm intra})$, dashed curve: 
amplitude of LM with the intra-orbital and inter-orbital charge-charge 
correlations $\langle \bm{S}^2 \rangle({\rm intra+cc})$, 
solid curve: amplitude of LM 
 with full correlations including the inter-orbital spin-spin correlations 
$\langle \bm{S}^2 \rangle({\rm total})$.
}
\label{figs2}
\end{figure}

Figure \ref{figs2} shows a systematic change of calculated amplitudes
of local moment (LM) from Sc to Cu.
The amplitude in the Hartree-Fock approximation shows a parabolic
behavior with the maximum 1.84 at $n_{e}=7$ (Mn).
When electron correlations are introduced, the amplitudes are enhanced.
The enhancement becomes larger near the half filling of $d$ electrons,
and amounts to 50 \% for Mn and Fe.

We have examined the role of electron correlations on the amplitude of
LM by adding three kinds of correlations successively.
We considered the amplitude with the intra-orbital correlations 
$\langle \bm{S}^2 \rangle({\rm intra})$, the amplitude with both the
intra-orbital and inter-orbital charge-charge correlations 
$\langle \bm{S}^2 \rangle({\rm intra+cc})$, 
and the amplitude with full
correlations $\langle \bm{S}^2 \rangle({\rm total})$ as shown in
Fig. \ref{figs2}. 
The intra-orbital correlations significantly enhance the amplitudes
because the correlations produce more active spins on each orbital with
suppression of the double occupancy.  The inter-orbital charge-charge
correlations do not make any significant contribution (see the
difference between $\langle \bm{S}^2 \rangle({\rm intra+cc})$ and 
$\langle \bm{S}^2 \rangle({\rm intra})$).  The
inter-orbital spin-spin correlations (,{\it i.e.}, the Hund-rule
correlations) make the active spins on different orbitals 
parallel to
each other, thus enhance further the amplitudes as shown in
Fig. \ref{figs2} (see the difference between 
$\langle \bm{S}^2 \rangle({\rm total})$ and 
$\langle \bm{S}^2 \rangle({\rm intra+cc})$). 
%
%
%
%

We summarize in Table \ref{tbls2} the calculated amplitudes of LM.
We find large amplitudes of LM, $\langle \bm{S}^2 \rangle = $ 2.87 and
2.58 for fcc Mn and bcc Fe, respectively.
In the case of Fe, the experimental value can be estimated from the 
observed effective Bohr magneton number~\cite{fallot44} 
$p_{\rm{eff}}\,(=3.20)$, 
because the Rhodes-Wohlfarth ratio is equal to 1.0 within $5\%$ error. 
Using the approximate relation 
$\langle \bm{S}^2\rangle=p_{\rm{eff}}^{2}/4$, we find the experimental
value $\langle \bm{S}^2\rangle = $ 2.56, being in good agreement 
with the present result $\langle \bm{S}^2\rangle=2.58$.

Alternative way of estimating the amplitude of LM is to use
the $3s$ inner core multiplet data in photoemission spectroscopy.
The multiplet splitting is expected even in the metallic system when the
$d$-electron charge fluctuations are significantly
suppressed~\cite{kake85}.  Using the experimental data of the $3s$ core
spectra and the localized model, the amplitudes of LM are
estimated to be $\langle \bm{S}^2\rangle = $ 2.8 for
$\alpha$-Mn~\cite{mcfee74} and $\langle \bm{S}^2\rangle = $ 2.5 for bcc
Fe~\cite{joyner80}.  These values are again in good agreement with the
present results 2.87 for fcc Mn and 2.58 for bcc Fe.

The amplitudes of LM calculated by the LA and the $d$ band
model are reported to be 4.0 for Mn and 2.9 for Fe~\cite{capell86}.  
These values are
considerably overestimated as compared with the present results: 
2.87 for Mn and 2.58 for Fe.
We suggest that the overestimate of the amplitudes is mainly due to the
neglect of the hybridization between the $sp$ and $d$ electrons 
in the LA + $d$ band model.
%
%
\begin{table}
\caption{
Calculated amplitudes of local moments $\langle \bm{S}^2 \rangle$
in iron-group transition metals.  
}
\label{tbls2}
\vspace*{5mm}
\begin{tabular}{ccccccccc}
\hline
Sc  & Ti  & V  & Cr  & Mn  & Fe  & Co  & Ni  & Cu \\ \hline 
1.148 & 1.809 & 2.284 & 2.407 & 2.868 & 2.583 & 1.733 & 0.895 & 0.260 \\ \hline
\end{tabular}
\end{table}
%
%

\section{Summary}

We have investigated the quantitative aspects of the ground-state
properties of iron-group transition metals from Sc to Cu on
the basis of the first-principles momentum dependent local ansatz
approach (MLA) which we recently developed.  The theory reduces to the
Rayleigh-Schr\"{o}dinger perturbation theory in the weak Coulomb
interaction limit, and describes quantitatively the ground-state of
correlated electrons by means of the self-consistent 
momentum-dependent variational parameters for the two-particle 
excited states.

We obtained the correlation energy, the local charge fluctuations, and
the amplitudes of local moments (LM) in the paramagnetic state, 
and clarified the role of three types of electron correlations in these
quantities: the intra-orbital correlations,
the inter-orbital charge-charge correlations, and the inter-orbital
spin-spin correlations (, {\it i.e.}, the Hund-rule correlations).

Calculated correlation energy gain curve 
shows a peak near the half-filled $d$ electron
number. We obtained large correlation energy gains 0.090 Ry for fcc Mn 
and 0.094 Ry for bcc Fe.  We found that both the intra-orbital and 
the inter-orbital correlations make significant contribution to the
correlation energies for Mn and Fe.  
The inter-orbital charge-charge correlations become 
significant when the $d$ electron number deviates from the
half-filling.

We calculated the Hund-rule coupling energies $\Delta \epsilon_{\rm H}$ 
which are not taken into account in the paramagnetic 
calculations of the DFT+LDA band theory.  
We found that the energy $\Delta \epsilon_{\rm H}$ shows a large value 
for Mn, Fe, and Co: 0.29 eV (Mn), 0.30 eV (Fe), 0.14 eV (Co), 
while it is not essential for the magnetism of Ni: 0.03 eV (Ni).
We pointed out that the Hund-rule coupling energy can resolve a large
difference in magnetic energy $E_{\rm mag}$ between the DFT+LDA and the
first-principles DCPA. 
The magnetic energy in the DFT+LDA, {\it i.e.}, $E_{\rm mag}({\rm LDA})$ 
is overestimated because of the lack of the Hund-rule coupling energy
$\Delta \epsilon_{\rm H}$ in the paramagnetic state.  Subtracting 
$\Delta \epsilon_{\rm H}$ from $E_{\rm mag}({\rm LDA})$, we obtained
$E_{\rm mag} \sim$ 2000 K for bcc Fe and 2600 K for fcc Co, 
being comparable to those obtained by the first-principles DCPA. 

The intra-orbital correlations suppress the double occupancy on the same
orbital to reduce the Coulomb energy on each orbital, so that the 
correlations
suppress the $d$ electron hopping and thus charge fluctuations.  
The inter-orbital charge-charge
correlations suppress the creation of electron pairs on the different
$d$ orbitals.  The correlations therefore also suppress the $d$ electron
hopping and lead to the suppression of charge fluctuations for $d$
electrons.  We found
that Mn shows the strongest suppression of charge fluctuations.  
The deviation from the Hartree-Fock charge fluctuations is found 
to be 0.92 in the case of fcc Mn.  
Calculated charge fluctuations are nearly constant from V to Fe due to
electron correlations: 1.57 (V), 1.57 (Cr), 1.52 (Mn), 1.44 (Fe).  
These values are roughly twice as large as those obtained by 
the early calculations based on the LA + $d$ band model.    

The intra-orbital correlations also increase the active spins on each
$d$ orbital as the result of suppression of the double occupancy on 
the $d$ orbitals.  The inter-orbital spin-spin correlations, {\it i.e}, 
the Hund-rule correlations make these spins on different $d$ 
orbitals parallel, and 
lead to the enhancement of the amplitudes of LM.  The enhancement 
effects become maximum near the half-filling of $d$ electrons.  We
found that the enhancement amounts to 50 \% for Mn and Fe, 
so that we obtained 
$\langle \bm{S}^2\rangle = $ 2.87 (Mn) and 2.58 (Fe).  These results are
in good agreement with the experimental 
values, 2.8 ($\alpha$-Mn) estimated from the inner core photoemission
data and 2.56 estimated from the effective Bohr magneton
number.  Early calculations based on the LA + $d$ band model 
overestimate the amplitudes of LM.

In order to make these conclusions clearer, it is desirable to investigate  
the ground-state properties of the ferromagnetic Fe, Co, and Ni with
use of the first-principles spin-polarized MLA.

\begin{acknowledgment}


The present work is supported by a Grant-in-Aid for Scientific Research (25400404).
Numerical calculations have been partly carried out with use of the 
facilities of the Supercomputer Center, the Institute for Solid State 
Physics, the University of Tokyo.

\end{acknowledgment}



\end{document}